\documentclass[12pt]{article}

\usepackage{psfig}

\newlength{\figwidth}
\newcommand{\eabe} {\begin{eqnarray}}
\newcommand{\eaen} {\end{eqnarray}}
\newcommand{\eqbe} {\begin{equation}}
\newcommand{\eqen} {\end{equation}}
\newcommand{\mrm} {\mathrm}
\newcommand{\srm}[1] {_{\mathrm{#1}}}
\newcommand{\ol} {\overline}
\renewcommand{\ln} {\mrm {ln}}

\newcommand{\bibl}[5]
	{#1, {\it #2} {\bf #3} (#4) #5}

\newcommand{\pair}[1] {${\mrm {#1 \ol #1} }$}

\begin{document}

\begin{titlepage}
\begin{flushright}
 LU TP 98-22\\
 November 1998
\end{flushright}
\vspace{25mm}
\begin{center}
  \Large
  {\bf Investigations of Quark Fragmentation Universality} \\
  \vspace{12mm}
  \normalsize
  Patrik Ed\'en\footnote{e-mail patrik@thep.lu.se}\\
  Department of Theoretical Physics\\
  Lund University\\
\end{center}
\vspace{5cm} {\bf Abstract:} \\ 
We propose event cuts in deep inelastic scattering, suitable for an
examination of quark fragmentation universality. We compare the
current hemisphere of the Breit frame with hemispheres in $e^+e^-$
annihilation events, using Monte Carlo simulations. The agreement
between the two processes is improved after the suggested events
cuts. A method to study the scale evolution in quark hemispheres using
data from fixed energy $e^+e^-$ experiments is presented. This makes
it possible to use the high statistics from LEP1 also at scales below
the ${\mrm Z}^0$ mass. We also discuss observables which are sensitive to the
dynamics of regions closer to the remnant. The observables probes the
relatively clean region on the current side of the hardest emission in
the event, and can be used to distinguish between different
assumptions about remnant effects and mechanisms for the parton
evolution.

\end{titlepage}
\section{Introduction}\label{sec:intro}
In deep inelastic scattering (DIS), the current hemisphere of the Breit frame is expected to be very similar to one hemisphere in an $e^+e^-$ experiment. This expectation relies on the fundamental assumption of quark fragmentation universality. 
However, a set of features in DIS introduce corrections to this assumption.

QCD radiation can give rise to high-$p_\perp$ emissions, where $p_\perp$$\sim Q$ or higher. These emissions have no correspondance in an $e^+e^-$ event, where the kinematical constraint for gluon emission is $p_\perp\le \sqrt s/2$. A high-$p_\perp$ emission in DIS significantly changes the kinematical properties of the struck quark, and a high-$p_\perp$ event sometimes manifests itself with a completely empty current region~\cite{empty}. If the rate of such events is getting non-negligeble, they may introduce an uncertainty to the interpretation of the data. Furthermore, problems arise even if the current region is not empty, but merely depopulated. It is therefore of interest to find a way to exclude high-$p_\perp$ events using another signal than an empty current Breit hemisphere. One candidate method is to reconstruct the highest $p_\perp$-scale of the event using jets, and we will in this paper discuss that approach.

The flavour composition in $e^+e^-$ and DIS is not exactly the same. When excluding high-$p_\perp$ events from the DIS analysis, the boson-gluon fusion channel for heavy quark production is supressed.
This implies a lower heavy quark rate in the studied DIS sample, as compared to $e^+e^-$ data at corresponding energies. A uds enriched $e^+e^-$ data sample with high statistics is available from the ${\mrm Z}^0$ peak.
 We will in this paper discuss a method to study properties of $e^+e^-$ quark hemispheres at different scales, using data from the fixed energy ${\mrm Z}^0$ experiments.

At very low squared momentum transfer, when $Q^2$ is of the same order as the squared $\rho$-meson mass $m_\rho^2$, the vector meson component of the photon affects the current region properties. We will restrict the discussion to $Q^2>4\mrm{GeV}^2\sim 8m_\rho^2$. In this kinematical range the vector meson component of the probe is suppressed, and it will not be discussed further in this paper.

Comparisons of DIS and $e^+e^-$ data support quark fragmentation universality~\cite{QFUres}. Some differences are however observed, especially at low $Q^2$. These can in general be qualitatively explained from the features discussed above, but data on e.g.\ strangeness rates in DIS are less well understood~\cite{s_anomaly}.

In this paper, we investigate how the similarity between $e^+e^-$ and current Breit hemispheres can be improved with proper event cuts. The effects of event cuts are examined with MC simulations. 
We also discuss observables  which are sensitive to the dynamics of regions closer to the remnant, where the theoretical situation is not so clear, due to e.g.\ the uncertainties from the BFKL Pomeron and from remnant effects. The observables discussed here are defined to probe the current side of the hardest emission in the event. This is a relatively clean region with a reduced dependence on soft remnant properties, as compared to the forward target region. This facilitates the interpretation of the results, which also can be compared to corresponding analyses in $e^+e^-$ experiments.
We show that the proposed observables can distinguish between different existing models.

The outline of this paper is as follows. In section~\ref{sec:jets} we discuss suitable jet algorithms in a search for high-$p_{\perp}$ events. In section~\ref{sec:CBHres}, the current region properties after different event cuts are studied and compared to $e^+e^-$ results. In section~\ref{sec:eeuds} we discuss a method to obtain experimental $e^+e^-$ data whith a controlled rate of heavy flavours. In section~\ref{sec:remnant} we propose observables which are sensitive to the dynamics of regions closer to the remnant. Finally, the results of this paper are summarized in section~\ref{sec:summary}.

\section{Jet Algorithms}\label{sec:jets}
In DIS events, high-$p_\perp$ emissions can break the quark fragmentation universality. In this section we discuss jet cluster algorithms suitable for a reconstruction of the highest $p_\perp$-scale of an event. The result may depend on the choice of jet finding algorithm used, and we will here discuss the properties of a set of possible algorithms.

Since the aim of the jet clustering is to find the highest $p_\perp$, it is natural to use $k_\perp$-type cluster algorithms. Our approach will be to combine jets in the order specified by a $k_\perp$ distance, until only three remain (including the remnant jet). The event is accepted if the $p_\perp$ specified by the final three jets is lower than $Q/2$. This cut is in analogy to the kinematical constraint for gluon emissions in $e^+e^-$, which is $p_{\perp{\mrm g}}\le E\srm{g}\le\sqrt{s}/2$.

Historically,  $k_\perp$ algorithms were first designed for $e^+e^-$ physics. A set of different algorithms exist~\cite{jetalgs}. In DIS, extra requirements are in general imposed on the cluster algorithm, since jet observables in DIS are dependent on structure functions. In a comparison with analytical calculations, a jet algorithm where the properties factorizes into perturbatively calculable coefficients convoluted with the structure functions is preferred. This implies that the remnant jet cannot always be treated on equal footing with the other jets. A $k_\perp$ algorithm designed to fulfill jet requirements in DIS is presented in~\cite{ktalg}.  However, in the present study we are merely interested in jets with relatively high $p_\perp$, and the infrared properties of the used algorithm are less important. Furthermore, we do not intend to investigate jet cross sections or the jets themselves, but merely to exclude high-$p_{\perp}$ events. In the accepted sample, the analysis is performed on the current Breit hemisphere, which is defined independently of any jets. We have therefore chosen to use the simpler $e^+e^-$ $k_\perp$ algorithms.

To estimate the reliability of the results obtained using jets, we have used three different algorithms, \textsc{Luclus}~\cite{jetset}, \textsc{Durham}~\cite{durham} and \textsc{Diclus}~\cite{diclus}, applied in the hadronic CMS system.
All these algorithms start off with the final state particles as initial clusters. These clusters are then iteratively merged in an order specified by a distance of $k_\perp$-type, until all remaining distances are larger than a resolution parameter, or until a specified number of clusters remain.

In the HERA experiments, particles which in the lab frame have  a large pseudo-rapidity w.r.t.\ the proton direction are not detected. 
In our analysis, we have chosen to exclude all particles with pseudo-rapidity larger than $3.8$, to take this into account.
For clustering purposes, an initial cluster is introduced along the proton direction, carrying the missing longitudinal momentum. (To exclude the possibility that  the obtained results are merely a consequence of the undetected target region, we have also performed analyses where all particles are used in the clustering. The result is that the effects presented in the paper would be even more prominent, if having access to a full coverage detector.)

In the \textsc{Durham} algorithm, the distance between the clusters $i$ and $j$ is defined as
\eqbe d_{ij}^{(D)2} = 4\min(E_i^2,E_j^2)\sin^2(\theta_{ij}/2). \eqen
The cluster pair with smallest distance value are combined into one new cluster, with a four-momentum given by the sum of its constituents. For small angles $\theta_{ij}$, the \textsc{Durham} distance $d_{ij}$ is the transverse energy of the softer particle with respect to the other. The factor $2\sin(\theta_{ij}/2)$, used instead of  $\sin(\theta_{ij})$, prevents unreasonable merging of oppositely moving clusters.

The \textsc{Luclus} algorithm is very similar, but uses the distance
\eqbe d_{ij}^{(L)2} = 4\frac{|{\mathbf p}_i|^2|{\mathbf p}_j|^2}{(|{\mathbf p}_i|+|{\mathbf p}_j|)^2}\sin^2(\theta_{ij}/2). \eqen
Apart from the difference between $|{\mathbf p}|$ and $E$, the \textsc{Durham} factor $\min(E_i,E_j)$ corresponds in the \textsc{Luclus} measure to $E_iE_j/(E_i+E_j)$. This implies that $d_{ij}^{(L)}$ always is somewhat smaller than  $d_{ij}^{(D)}$. To the \textsc{Luclus} algorithm, a reassignment scheme is added, where all particles are moved to the closest existing cluster after each merging. At the very end of the algorithm, the reassignment procedure is iterated til convergence. This scheme prevents particles to, through successive mergings, end up in a jet moving in a very different direction. 

A cluster algorithm can be viewed as a reconstruction of a parton cascade backwards. The two described algorithms combines two clusters into one, like an inverse parton emission. In the \textsc{Diclus} algorithm, supplied with the \textsc{Ariadne} MC~\cite{ariadne}, three clusters are merged into two. This corresponds to an inverse dipole emission. The distance measure used is
\eqbe \left[d_{i(jk)}^{(A)}\right]^2 = \frac{\left[s_{ij}-(m_i+m_j)^2\right]\left[s_{ik}-(m_i+m_k)^2\right]}{s_{ijk}}. \eqen
Here $s_{ij}$ and $s_{ijk}$ are the squared masses of the clusters indicated by the indices. This distance is a Lorentz invariant $p_\perp^2$-measure for parton $i$ with respect to a dipole built up by $j$ and $k$. In the merging, the cluster $i$ is absorbed into $j$ and $k$. The momentum of $i$ is distributed in accordance with the recoil treatment in the dipole cascade~\cite{ariadne}.

In the present analysis, three clusters are constructed using one of the presented algorithms. These final clusters are called jets, and the smallest remaining $d_{ij}$ specifies the $p_\perp$ of the event.

The three chosen algorithms are all of $k_\perp$--type, and hence similar in many respects. For quasi-collinear, massless clusters, the three distance measures are very similar. There are however also noticeable differences. Differences in the distance measures occur at large angles and for heavy clusters, the reassignment procedure in \textsc{Luclus} is not used by the other algorithms, and the merging of three clusters into two is a unique feature of the \textsc{Diclus} algorithm.
In the following, we will investigate results using all three algorithms, and we will consider results which are insensitive to the specific choice of algorithm as reliable.

\begin{figure}[tb]
  \begin{center}  \hbox{ \vbox{
	\mbox{\psfig{figure=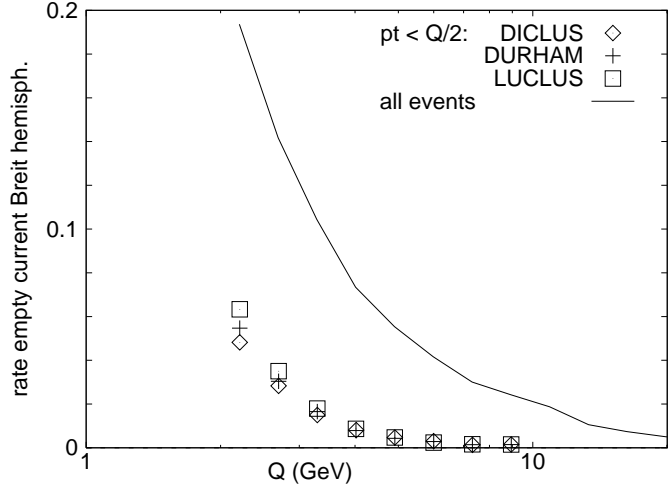,width=\figwidth}}
   }  }
  \end{center}  \caption{\em The rate of empty current Breit hemispheres. Results from MC simulations. The solid line shows the rate when considering all generated DIS events. The symbols show the result when events with a jet-$p_\perp> Q/2$ are excluded, using three different $k_\perp$ clustering algorithms. The rate is significantly reduced after the cut in jet-$p_\perp$.}
  \label{f:empty}
\end{figure}

\section{Current Breit Hemisphere Properties}\label{sec:CBHres}
To examine if a cut in jet-$p_\perp$ is suitable to isolate a sample where universality is expected to hold, we will here study results for the current Breit hemisphere in MC generated events at HERA energies. The results are compared to $e^+e^-$ MC results.

We simulate the electro-weak interaction in DIS  using the \textsc{Lepto} MC~\cite{lepto}. In both the $e^+e^-$ and DIS simulations, we use the Colour Dipole Model~\cite{CDM}, implemented in \textsc{Ariadne}~\cite{ariadne}, to describe the parton cascade. The  Lund string fragmentation model~\cite{stringmodel}, implemented in \textsc{Jetset}~\cite{jetset} is then used to describe hadronization. The chosen MC programs give a generally good description of data in DIS an $e^+e^-$.

\subsubsection*{Effects of the $p_\perp$-cut}
In Fig~\ref{f:empty}, the rate of events in DIS with empty current Breit hemispheres is shown to be significantly reduced after a $p_\perp$$<Q/2$ cut, for all jet finding algorithms. As discussed in section~\ref{sec:jets}, the \textsc{Luclus} distance measure is less restrictive than the \textsc{Durham} one, and this is seen as a slightly higher rate of empty events. The rate is however significantly reduced with all three algorithms.

Though the phase space for the current Breit hemisphere is very similar to an $e^+e^-$ hemisphere, there are differences in the regions nearby, which affect the average energies. Consider two hemispheres in the rest frame of an $e^+e^-$ annihilation event. In general, they have different masses and hence different energies. In other words, a high-$p_\perp$ emission in one hemisphere reduces the energy of the other, due to energy--momentum conservation.

As for an $e^+e^-$ hemisphere, the energy of the current Breit hemisphere at fixed $Q^2$ is expected to deviate from $Q/2$. The energy depends on emissions in this hemisphere and in the nearby phase space of the opposite hemisphere. The latter is in DIS very large, but after our suggested event cuts, it is reduced by the condition $p_\perp$$<Q/2$. The corresponding kinematical constraint in $e^+e^-$ is however $\left| \mathbf{p}\right|<Q/2$, which is more restrictive. Thus the region for high-$p_\perp$ emissions which can reduce the energy of the considered hemisphere is larger in DIS than in $e^+e^-$. The emission density at this relatively high $p_\perp$ is suppressed, but not zero. Without reaching for a quantitative prediction, we may conclude that we expect the mean energy and multiplicity of the current Breit hemisphere to satisfy
\eqbe \left<E\srm{CBH}\right> < \frac Q 2,~~~\left<N\srm{CBH}\right> < \frac1 2N_{ee}(Q^2), \eqen
where $N_{ee}(Q^2)$ is the average multiplicity in an $e^+e^-$ experiment with invariant mass $Q^2$.

\begin{figure}[tb]
  \begin{center}  \hbox{
     \vbox{
	\mbox{
	\psfig{figure=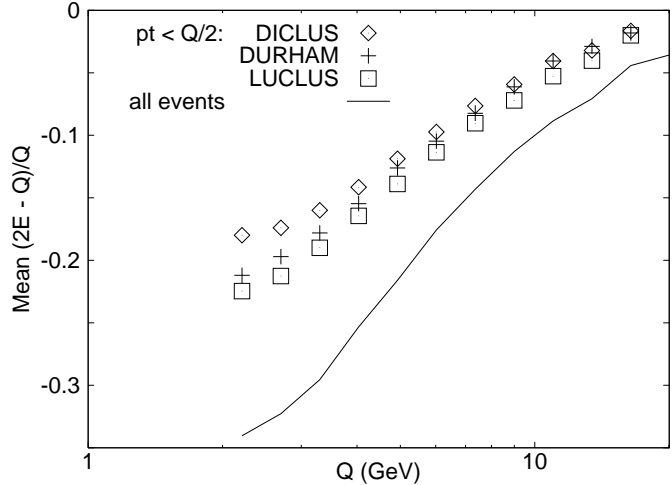,width=0.55\textwidth}
	}
    }
  }
  \end{center}  \caption{\em The relative difference between average energy and $Q/2$ in the current Breit hemisphere. The relative shift in the MC simulations is significantly reduced after a $p_\perp$ $<Q/2$ cut.}
  \label{f:Ediff}
\end{figure}
If $2\left<E\srm{CBH}\right>/Q$ would be much smaller than 1, the similarity between the current Breit hemisphere and an unbiased $e^+e^-$ hemisphere is poor, and it is not likely that reliable conclusions from a comparison can be drawn. On the other hand, if $2\left<E\srm{CBH}\right>/Q \approx 1$, the current Breit hemisphere sample may be closely related to an unbiased $e^+e^-$ hemisphere.
In Fig~\ref{f:Ediff}, the relative energy shift $2\left<E\srm{CBH}\right>/Q - 1$ is shown to be sizeable for the unrestricted event sample, but significantly reduced after imposing a $p_\perp < Q/2$ cut
. 
\subsubsection*{Flavour Compositions}
In a comparison between $e^+e^-$ and DIS, also the flavour composition in the different experiments need to be considered. In $e^+e^-$ the rate of heavy flavours varies with energy. On the ${\mrm Z}^0$ pole it is determined by the couplings to the ${\mrm Z}^0$, while at lower energies it is determined by the quark electric charges. 
In DIS events at HERA energies, photon exchange is dominating over ${\mrm Z}^0$ exchange. However, here the flavour composition also depends on soft physics. The rate of charm in the proton structure function (intrinsic charm) is suppressed by the heavy c mass. 

A cut in $p_\perp$ suppresses the boson-gluon fusion channel for charm production in DIS. This implies that the c rate in the low-$p_\perp$ sample is reduced. Using MC simulations, we have found charm to be suppressed  approximately by a factor 2 in a $p_\perp< Q/2$ event sample, with all cluster algorithms.
A problem for the interpretation of this result is the fact that the MC simulation program significantly underestimates the rate of charm events. In the MC this is about 10\%, while data indicate a much larger value around 25\%~\cite{charmrates}.
However, a similar suppression ought to be expected also for real data. If  boson-gluon fusion is a more important source of charm in Nature than assumed in the MC model, the suppression may be even stronger.  It would then be appropriate to compare with uds enriched $e^+e^-$ data. How to obtain such data at different energies will be discussed in section~\ref{sec:eeuds}.

\begin{figure}[tb]
  \begin{center}  \hbox{ \vbox{
	\mbox{\psfig{figure=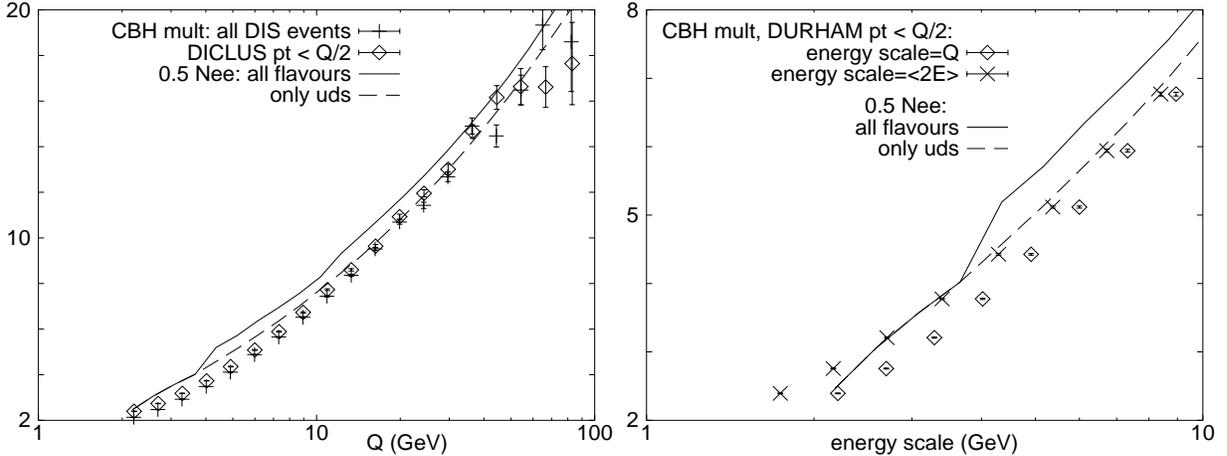,width=0.98\textwidth}}
   }  }
  \end{center}  \caption{\em MC simulations of multiplicities in $e^+e^-$ and the current Breit hemisphere (CBH). {\bf LEFT:} $e^+e^-$ results with squared mass $s=Q^2$, allowing for all flavours (solid line) and only uds (dashed line), compared to  CBH multiplicities using all events (crosses) and a jet-$p_\perp < Q/2$ sample (diamonds). Both the exclusion of high-$p_{\perp}$ DIS events and heavy flavour $e^+e^-$ events improves the agreement between the multiplicities. {\bf RIGHT:} For low energy scales, the agreement is further improved if comparing the CBH multiplicity with $e^+e^-$ data with squared mass $s=4\left<E\srm{CBH}\right>^2$.}
  \label{f:multuds}
\end{figure} 
Monte Carlo results for average multiplicities are presented in the left plot of Fig~\ref{f:multuds}. The results for the  low-$p_\perp$ DIS sample are essentially equivalent for the three considered jet algorithms, and only the \textsc{Diclus} result is shown. 
In this MC generated event sample, heavy quark rates are well below 10\%, and it is therefore better compared to a MC $e^+e^-$ sample with only uds. 
The agreement between the low-$p_{\perp}$ and the $e^+e^-$ uds sample is better than between the unrestricted event samples also shown in Fig~\ref{f:multuds}. 

\subsubsection*{Energy Scale Corrections}
As discussed previously in this section, high-$p_\perp$ emissions close in rapidity to the considered hemisphere implies a reduction of the average energy in the current Breit hemisphere, $\left<E\srm{CBH}\right>$, to a value slightly smaller than $Q/2$. It is then natural to compare the current Breit hemisphere multiplicities with $e^+e^-$ data at a squared mass $s=4\left<E\srm{CBH}\right>^2$. The agreement between $e^+e^-$ and Breit frame multiplicities is then significantly improved, as shown in the right plot of Fig~\ref{f:multuds}.  Only the \textsc{Durham} result is presented, but the other results are very similar.

When compensating for small systematic differences by changing scale to the average energy, many effects -- known as well as unknown -- may be corrected for. However, after correcting for expected differences in this way, unexpected features which break quark fragmentation universality can still be searched for in observables less blunt than the mean multiplicity, like  strangeness rates, energy spectra and higher multiplicity moments.

Comparing the left and right plot in Fig~\ref{f:multuds}, we see that the high-$p_\perp$ cut has a relatively small influence on the average multiplicity, as compared to the energy scale shift. However, in Fig~\ref{f:empty} we note that the rate of empty current Breit hemisphere events, which are excluded from the analysis more by necessity than by theoretical understanding, are reduced with the high-$p_\perp$ cut. This is also the case for the energy shift, as seen in Fig~\ref{f:Ediff}. Thus we find that a cut in jet-$p_\perp$ is a powerful step towards an event sample where quark fragmentation universality is expected to hold.

\section{Scale Evolutions in Fixed Energy $\mathbf{e^+e^-}$ Annihilation}\label{sec:eeuds}
A large sample of uds enriched events are available from LEP1 at $\sqrt s=90$GeV. This is not the case for other energies corresponding to the HERA kinematical range.
In this section we discuss how the scale evolution of $e^+e^-$ uds hemispheres can be examined, using data from a fixed energy $e^+e^-$ experiment.

In search for an algorithm which enables such a study, it is useful to recollect the Breit frame analysis discussed previously in this paper. In that analysis three jets are found with a $k_\perp$ algorithm in the hadronic center of mass frame. In events were the $p_\perp$ of the event is smaller than $Q/2$, the current Breit hemisphere is investigated. This region corresponds to a cone in the hadronic center of mass frame, defined by a constraint on the rapidity measured in the current direction,
\eqbe y>\frac1 2 \ln(W^2/Q^2), \label{e:CBHcone} \eqen
where $W^2$ is the squared mass of the hadronic system.

We will now consider a very similar algorithm for $e^+e^-$. In an $e^+e^-$ experiment at squared mass $s$, we consider an artificial scale $Q\srm{max}^2<s$. Three jets are reconstructed with a $k_\perp$ cluster algorithm, and events where $p_\perp$$> Q\srm{max}/2$ are excluded. In the accepted event sample,  $Q\srm{max}/2$ can be identified with the maximal allowed virtuality for a parton in the event. Thus the multiplicity depend on two scales, energy and virtuality. For $Q\srm{max}^2=s$, the two scales coincide and we have an unbiased event sample.

We study the multiplicity for particles where the rapidity, measured in the thrust direction, satisfies
\eqbe \left|y\right|>\frac1 2 \ln(s/Q\srm{max}^2). \label{e:tcone} \eqen  
In $e^+e^-$, where no proton remnant is present, both the positive and negative rapidity cone is considered. These cones in the thrust direction correspond to $e^+e^-$ hemispheres with squared invariant mass $Q\srm{max}^2$, and their evolution with this scale can thus be studied in a fixed energy experiment. (For a more detailed discussion, see e.g.~\cite{jetscales}.)

The ``thrust-cone'' method presented here is designed to be as similar as possible to the Breit frame analysis. In both $e^+e^-$ and DIS, jets reconstructed in the hadronic rest frame are used to exclude high-$p_\perp$ events. In the accepted events, the analysis is performed in rapidity cones defined by Eq~(\ref{e:CBHcone}) and Eq~(\ref{e:tcone}), respectively. These cones correspond in some other Lorentz frame to unbiased hemispheres, where the maximal allowed transverse momentum coincides with the hemisphere energy. (In DIS, the boost to the relevant Lorentz frame, the Breit frame, is performed explicitly.)

Two differences are present due to the absence of a $t$-channel probe in $e^+e^-$. The scale $Q^2$ and the direction which defines the rapidity cones are determined by the probe in DIS. In  $e^+e^-$, the scale $Q\srm{max}^2$ is chosen freely and the rapidity is measured along the thrust direction.

The upper limit on $p_\perp^2$ in a jet corresponds to the maximal allowed virtuality for the parton initiating the jet. Jet definitions suitable for a systematic study of the dependence upon the two separate scales, energy and virtuality, are presented in~\cite{jetscales}, and will be further discussed in the section~\ref{sec:remnant} of this paper. In the comparison of $e^+e^-$ results with data from the current Breit hemisphere in DIS, we are however mainly interested in quark jets where the two scales coincide. It is then sufficient to study the scale evolution using the relatively simple thrust-cone method.

\begin{figure}[tb]
  \begin{center}  \hbox{
     \vbox{
	\mbox{ \psfig{figure=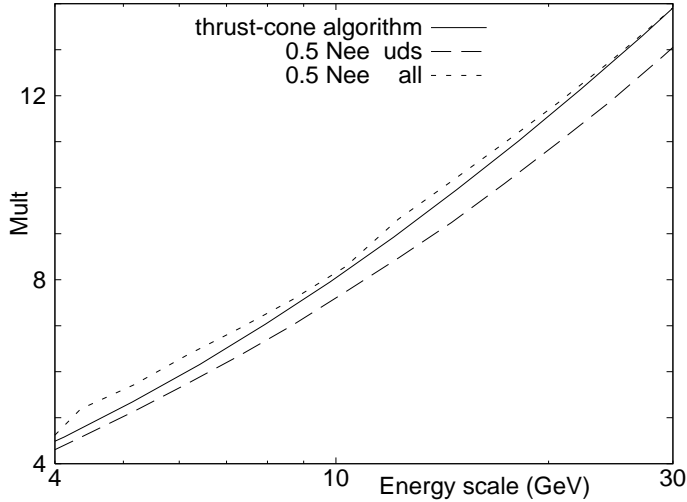,width=\figwidth}	}
    }
  }
  \end{center}  \caption{\em Multiplicities in MC simulated $e^+e^-$ events including all allowed flavours (dotted line) are higher than for light quarks events (dashed line). The multiplicity obtained by the thrust-cone algorithm in a uds sample at the ${\mrm Z}^0$ pole (solid line) are in better agreement with the uds result for moderate scales between 4 and 8 GeV. This corresponds to current Breit hemispheres in DIS with $Q^2$ between 15 and 60 $\mrm{GeV}^2$. For larger scales the thrust-cone results differ from the uds sample in a similar way as the full flavour results.}
  \label{f:eescales}
\end{figure}
In Fig~\ref{f:eescales} the multiplicity evolution of an $e^+e^-$ uds sample is compared to results including all flavours and results obtained by the thrust-cone algorithm in a fixed energy uds sample. The thrust-cone multiplicities are plotted as a function of the average energy scales obtained in the cones. These scales are somewhat smaller than the chosen $Q\srm{max}/2$, and the deviations are of order  5 to 10\%.

In rare cases where an emitted gluon gives the hardest jet, the thrust will be directed along the gluon momentum. Thus the average multiplicity in the thrust-cone analysis will get some contribution from gluon jets, which have a larger multiplicity.  
This results in a systematic overestimation of the multiplicity, as shown in Fig~\ref{f:eescales}. 

The aim of the thrust-cone algorithm is to reproduce the scale evolution of a $e^+e^-$ uds sample  better than a sample including all flavours. As seen in Fig~\ref{f:eescales}, this is achieved for moderate $Q\srm{max}$, between 4 and 8 GeV. This corresponds to current Breit hemispheres in DIS with $Q^2$ between 15 and 60 $\mrm{GeV}^2$, which is an important range at the HERA experiments. 

For $Q\srm{max}$ larger than 8 GeV, where large $p_\perp$ gluon jets are not excluded, the effect of gluon ``pollution'' is of the same order as the effect of the heavy quarks. 
However, we note that Fig~\ref{f:eescales} shows results from MC simulations. 
In a realistic experimental analysis, the thrust-cone results applied on uds events at the ${\mrm Z}^0$ peak will benefit from very large statistics compared to experiments at lower energies, and also from the fact that the scale evolution can be studied over a large range with the same detector. To conclude this section, our investigation indicates that it would be interesting to compare DIS data not only to full $e^+e^-$ results at different $s$, but also with thrust-cone results in uds samples from the ${\mrm Z}^0$ experiments.

\section{Approaching the Proton Remnant}\label{sec:remnant}
Compared to the current Breit hemisphere, the target region in the Breit frame is theoretically much less understood. It is difficult to determine the basic dynamics (e.g.\ the importance of the BFKL mechanism) using only inclusive observables, like $E_\perp$-flow. In the far forward direction of the target region, results depend to a large extent on non-trivial soft properties of the remnant. This region is also more difficult to study in many experiments. E.g.\ at HERA, many particles are lost in the beam pipe.

In this section we discuss two observables which examine the current and central rapidity ranges, but which nevertheless are sensitive to soft remnant effects and the dynamics of the evolution. The first observable is actually the same one that has been discussed throughout the paper, i.e.\ the multiplicity in the current Breit hemisphere. We will however here focus on results for very high $Q^2$. The rapidity distance to the remnant is then reduced, and as we will see, different models for remnant effects predict different results in this observable.

We will also propose a method to gradually extend the investigated phase space region, from the current Breit hemisphere, to also include central rapidity regions of the target hemisphere. 
Compared to the forward target region, this region is less dependent on soft remnant effects, but the basic mechanism for radiation is not very well known and depends to some extent on rather subtle  coherence effects (cf.\ the ``Feynman--Gribov puzzle'' discussed in~\cite{FGpuzzle}).

We will use a jet definition, ``the Mercedes algorithm''~\cite{jetscales}, with which we can construct quark jets with well-defined rapidity regions both in DIS and $e^+e^-$. This algorithm can be used to examine current jets in DIS, where the considered rapidity range is large enough to reach into the target hemisphere of the Breit frame, and the result  can be compared to similarly defined jets in $e^+e^-$.

\begin{figure}[tb]
\begin{center}  \hbox{
     \vbox{
	\mbox{
	\psfig{figure=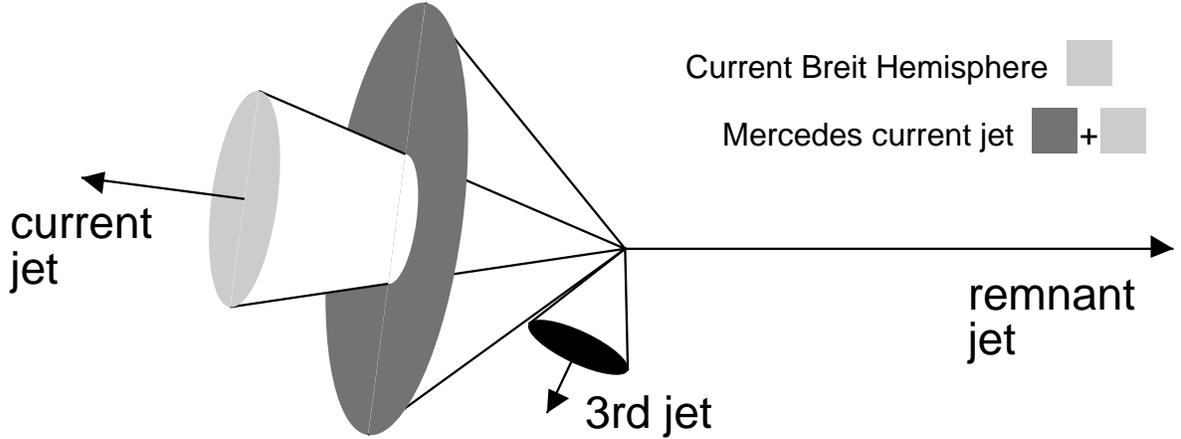,width=0.95\textwidth}
	}
    }
  }
  \end{center}  \caption{\em The current Breit hemisphere roughly corresponds to a cone around the current jet in the hadronic center of mass frame (light-gray cone). With the Mercedes algorithm, a current jet region defined to cover the rapidity range up to a resolved high-$p_\perp$ jet is studied (both gray cones). In the figure the third jet lies outside the current Breit hemisphere region and the Mercedes current jet will in this example include central rapidities in the target hemisphere of the Breit frame.}
  \label{f:cones}
\end{figure}
The Mercedes algorithm in DIS is applied to three-jet configurations, with  a current jet, a remnant jet and a third jet. Around the current jet a cone-like region is studied, which covers rapidities up to the third resolved jet.  If this is found outside the region which defines the current Breit hemisphere, as illustrated in Fig~\ref{f:cones}, the Mercedes current jet will probe central rapidities in the target hemisphere of the Breit frame. It will however only cover rapidities on the current side of the highest $p_\perp$ emission. This is a comparatively clean region in which relatively few dynamical mechanisms are important, as compared to more forward regions of the target hemisphere.

We will in this section start to present the Mercedes algorithm, applied $e^+e^-$ and DIS, respectively. The scale dependences and expected properties of the Mercedes jets are then discussed. After a presentation of some different models in DIS, we finally show that Mercedes jet results, and also the current Breit hemisphere results at high $Q^2$, can be used to distinguish between these.

\subsection{The Mercedes Algorithm} 
The starting point of the Mercedes algorithm is a three-jet configuration in $e^+e^-$, obtained by a $k_\perp$ cluster algorithm in the center of mass frame.
The $p_\perp$ of the $e^+e^-$ event is specified as the smallest cluster $k_\perp$-distance between the remaining jets. This  implies that no emissions within the jets have higher transverse momentum, and  $p^2_\perp$  can thus be identified with the maximal virtuality of a parton initiating a jet in the event. 

The jets are analysed after a boost to the symmetric ``Mercedes'' frame, where all jet angles are $\theta^\prime = 120^\circ$. As the jet definition is not Lorentz invariant, such a Lorentz transformation can transfer soft particles from one jet to another. Thus the mass of a jet is not invariant. Instead the jet 
direction can be interpreted as the direction of a massless parton initiating the jet, which implies that it transforms as a light-like vector. (The \textsc{Diclus} algorithm is a Lorentz-invariant cluster algorithm, which thus is an exception to this discussion. However, particles are not assigned to any specific jet in this algorithm, and the produced jets are explicitly massless.)

In the Mercedes frame the boundaries between the jets are determined by the bisectors between them. 
In~\cite{jetscales} it is shown that when the gluon is significantly softer than the quarks, the rapidity ranges (at a lower cut-off scale $\Lambda$) for the quark jets are given by
\eqbe y\srm q\approx\frac1 2\ln\left(\frac{4p^2_\perp}{\Lambda^2}\right)+\ln\left(\frac{1}{2(1-x\srm q)}\right). \label{e:yqrange} \eqen
The first term on the right-hand side of Eq~(\ref{e:yqrange}) is the rapidity range in the forward region of the jet where $2y>\ln(s/4p_\perp^2)$, which corresponds to an unbiased hemisphere of an event with squared invariant mass equal to $4p_\perp^2$ (cf.\ Eq~(\ref{e:tcone})). The second term is the extra rapidity range included in the Mercedes quark jet. In this range the maximal allowed transverse momentum is $p_\perp$. 

Thus the jet properties depend on two independent scales. The virtuality scale $p_\perp$ determines the size of the first range and the maximal allowed transverse momentum in the second, while the energy sets the size of the second range. This is not a unique feature of the Mercedes jet, all jets can in general be subdivided into two regions in a similar way, and thus depend on two scales. When looking at a forward cone, so that the second term vanishes, the two scales coincide. The cone then corresponds to an unbiased $e^+e^-$ hemisphere.
In the region corresponding to the second term in Eq~(\ref{e:yqrange}), where the maximal transverse momentum is constant, the emission density in an $e^+e^-$ jet is essentially flat in rapidity. This implies  that the multiplicity in the jet is expected to grow linearly with the allowed rapidity range, for fixed virtualities.

When applying the Mercedes algorithm to DIS events, we will use the \textsc{Durham} $k_\perp$ cluster algorithm in the hadronic center of mass frame, in order to resemble the corresponding $e^+e^-$ analysis. Three jets are reconstructed, and the smallest remaining $k_\perp$-distance between these specifies the $p_\perp$ of the event. The current jet is  tagged as the one with smallest polar angle to the probe direction, and is analyzed in the Mercedes frame. 
The considered rapidity range for the current jet is then given by Eq~(\ref{e:yqrange}), with $x\srm q$ being the  scaled energy $2E\srm q/W$ of the outgoing struck quark in the hadronic center of mass frame. 

The Mercedes current jet can reach into the target region of the Breit frame. The rapidity range for the current Breit hemisphere (the light-gray cone in Fig~\ref{f:cones}) is $\ln(Q/\Lambda)$.  Eq~(\ref{e:yqrange}) implies that the Mercedes current jet in DIS will cover a region which extends into the target hemisphere of the Breit frame (the dark-gray cone in Fig~\ref{f:cones}) when $p_\perp>Q(1-x\srm q)$.

As for the $e^+e^-$ case, the properties of the Mercedes current jet in DIS depends both on its energy and its virtuality. It may, however, also depend on the DIS scales  $x$ and $Q^2$. These scale dependencies can be investigated for a set of observables.  E.g., the Mercedes jet properties in events where $p_\perp$$>Q$ could be used to investigate the mechanisms behind resolved photon events~\cite{jung}. It is also possible to investigate different rapidity intervals in the Mercedes current jet separately.

Here we will investigate whether the Mercedes jet can be used to distinguish between different models in DIS. It is then sufficient to study the simplest possible observable, the average multiplicity in the whole jet. 
We will compare the rapidity range dependence of the average multiplicity in the DIS current jet and $e^+e^-$ quark jets, keeping the virtuality scale fixed.
We note that when $Q$ is smaller than $p_\perp$, a significant $Q^2$ dependence could be expected for the current jet in DIS. For the jets in the two experiments to be similar, we therefore exclude DIS events where $p_\perp > Q/2$. In the remaining event sample, we study the inclusive results, integrated over $Q^2$ and $x$.

\subsection{Models}
To demonstrate that the proposed observable can distinguish between different production mechanisms, we study some different models and see that they indeed give different results. We have chosen two versions of the \textsc{Lepto} MC~\cite{lepto}, based on a DGLAP parton evolution, and two versions of the \textsc{Ariadne} MC~\cite{ariadne}, implementing the soft radiation model~\cite{srm} with boson-gluon fusion included. All the chosen models uses the Lund string fragmentation to describe hadronization. The models however differ in the treatment of parton evolution and proton remnant effects.

In the \textsc{Lepto} MC, a parton shower based on the DGLAP evolution equation is implemented. This implies that parton emissions are strongly ordered in rapidity and transverse momentum. A ladder diagram describing a DIS event specifies a colour ordering for the emissions, which due to colour coherence  also specifies the order in rapidity. For large $Q^2$, the leading order contribution to the cross section is given by diagrams where the transverse momenta are strongly ordered, so that the parton closest in rapidity to the remnant has the smallest $p_\perp$, and  the parton closest to the quark interacting with the probe the largest. It is diagrams with this ordering in transverse momentum that are considered in the DGLAP evolution.

A model for soft colour interaction~\cite{SCI} (SCI) is supplied in the \textsc{Lepto} MC. In this model, the proton remnant may interact with the perturbatively produced partons by  exchanging soft colour quanta. These have negligible effect on the parton momenta, but alter the colour flow of the parton configuration. This implies that a \pair q pair produced in a boson-gluon fusion event may change its colour state from an octet to a singlet. This introduces a rapidity gap in the colour flow from the current to the remnant. In cluster and string fragmentation models the colour topology determines the phase space for hadron production, and thus the same rapidity gap manifests itself in the hadronic final state. This is a way to model diffractive DIS events, alternative to the Pomeron approach. 

Another possible consequence of SCI is that the colour flow is enlarged, thus producing more hadrons. This increases the $E_\perp$-flow in a way which is consistent with data~\cite{SCI}.

In the default version of the \textsc{Ariadne} MC, the radiation in DIS is expressed in terms of a dipole spanned from the struck quark to the remnant. The emitted partons are not necessarily ordered in $k_\perp$ and rapidity as in the DGLAP case, but the fact that the remnant is an extended object, and not point-like, gives rise to a suppression of high-$p_\perp$ emissions in the remnant direction.  Though this model is rather different from the DGLAP formalism, the radiation from the struck quark in the current region is treated similarly in the two Monte Carlos.

The  suppression of high-$p_\perp$ emissions, in the way it is modeled in \textsc{Ariadne}, may reach into the current Breit hemisphere when $Q^2$ is very high. 
It could however be argued that the suppression in this case is overestimated, since the maximal allowed virtuality for the struck quark then is lower than $Q^2$. A modification has therefore been developed, where the unsuppressed phase space is enlarged in high $Q^2$ events, to always include the expected region for final state emissions from the struck quark~\cite{hiQariadne}.

\subsection{Results}
We will here study the prediction of different MC simulation models on two observables, the multiplicity in the current Breit hemisphere at high $Q^2$ and the current jet multiplicity obtained by the Mercedes algorithm. 

\begin{figure}[tb]
\begin{center}  \hbox{
     \vbox{
	\mbox{
	\psfig{figure=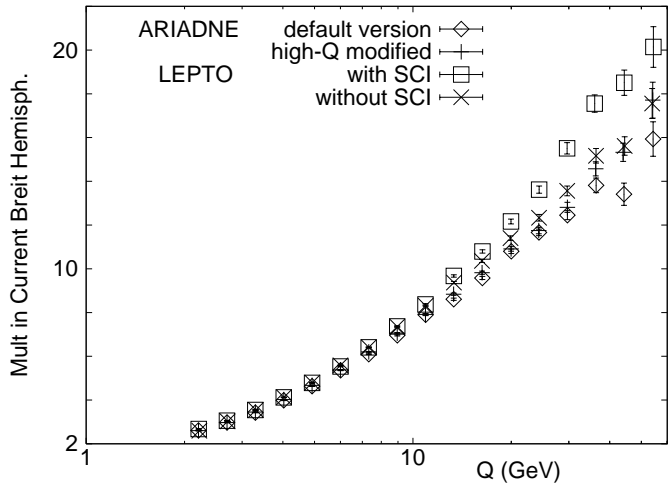,width=\figwidth}
	}
    }
  }
  \end{center}  \caption{\em Different model predictions for the multiplicity in the  current Breit hemisphere.  In the \textsc{Ariadne} default version the remnant suppression of high-$p_\perp$ emissions can reach into the current Breit hemisphere in events with high $Q^2$. This implies a lower multiplicity. In the \textsc{Lepto} MC with soft colour interaction, where the hadronization phase space in terms of a colour string can be enlarged, the multiplicities lie above the others.}
  \label{f:hiQ2}
\end{figure}
In  Fig~\ref{f:hiQ2}, the prediction for the multiplicities in the  current Breit hemisphere are shown, using different models. At very  high $Q^2$, the rapidity distance to the remnant is reduced, and different assumptions about remnant effects give different predictions
.

The differences can be understood from the remnant treatments of the models.
The \textsc{Ariadne} default version, where the remnant suppression of high-$p_\perp$ emissions can reach into the current Breit hemisphere for very large $Q$, predicts a lower multiplicity than the other models. The prediction from \textsc{Lepto} with soft colour interactions (SCI) lies above the others. 
SCI can give rise both to rapidity gaps and an extended colour flow, which increases the hadron multiplicity. The results here indicates that the latter effect dominates in the current rapidity region.

The \textsc{Lepto} MC without SCI and the high-$Q$ modified \textsc{Ariadne} MC differ mainly in the evolution of the parton distributions. In \textsc{Lepto} the evolution is based on the DGLAP equation, while the \textsc{Ariadne} MC implements a soft radiation model based on colour dipoles. The properties of the current Breit hemisphere are expected to be essentially independent of the form of the parton distribution evolution. The two MC results are similar at very high $Q^2$, in agreement with this expectation.

The MC result for an $e^+e^-$ uds sample (not shown) agrees well with the high-$Q$ modified \textsc{Ariadne} and \textsc{Lepto} without SCI.  Whether this agreement between $e^+e^-$ and DIS results with similar flavour compositions holds also for experimental data is however not fully investigated yet. The differences between the results in  Fig~\ref{f:hiQ2} show that the high-$Q^2$ Breit frame analysis can be used to discriminate between different models.

\begin{figure}[tb]
  \begin{center}  \hbox{
     \vbox{
	\mbox{
	\psfig{figure=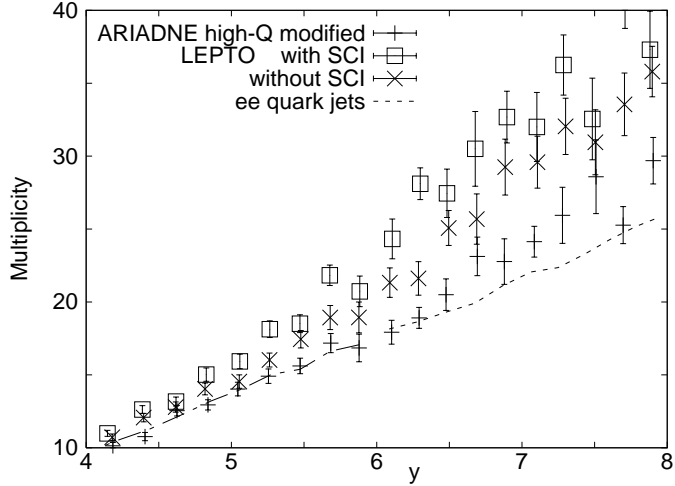,width=0.55\textwidth}
	}
    }
  }
  \end{center}  \caption{\em MC results on the multiplicity in DIS current jets and $e^+e^-$ quark jets using the Mercedes algorithm, as a function of the rapidity range in Eq~(\ref{e:yqrange}). The transverse momentum scale is constrained to 3 GeV $<p_\perp<$ 3.6 GeV. }
  \label{f:merc}
\end{figure}
Fig~\ref{f:merc} shows the Mercedes algorithm multiplicity of the current jet in DIS and quark jets in $e^+e^-$,  as a function of the available rapidity range $y$. The transverse momentum is constrained to the range 3GeV$<p_\perp<$3.6GeV and DIS events with $Q/2<p_\perp$ are excluded from the analysis. We have investigated the results for a rather low $p_{\perp}$, in order to make the analysis complementary to the high-$Q^2$ analysis shown in Fig~\ref{f:hiQ2}. The multiplicity of the $e^+e^-$ jets rise fairly linearly with $y$, as expected. For low values of $y$ the examined region lies within the current Breit hemisphere, and the DIS results are similar to the $e^+e^-$ result. 

Ideally, all jets correspond to a perturbative gluon emission, with similar $p_\perp$. However, ``junk jets'' may be formed, where  hadrons which happen to be close in phase space are merged into a resolved jet, even though no corresponding gluon was emitted. These ``junk jets'' are more easily formed when the multiplicity is high. Thus their presence in the considered event sample increases the average multiplicity. If the probability  to emit gluons giving rise to ``proper'' jets is reduced, the fraction of junk jet events in the accepted sample increases, and their effect on the result gets more prominent. The results for Mercedes jets with large rapidity ranges are based on events where one jet is found in the target region. There the probability for gluon emission is smaller than in a corresponding region in $e^+e^-$, and in Fig~\ref{f:merc} the junk jet contribution is seen as a larger multiplicity for DIS Mercedes jets with large rapidity ranges. 

The probability for jet-generating gluons in the target region differ in the considered DIS models, being lower for the DGLAP based \textsc{Lepto} MC than the dipole based \textsc{Ariadne} MC.
Thus the high-$Q$ modified \textsc{Ariadne} MC and the \textsc{Lepto} MC without SCI, which give very similar results in the high-$Q^2$ Breit frame analysis of Fig~\ref{f:hiQ2}, differ in the Mercedes algorithm results in Fig~\ref{f:merc}.
 This result suggests that properties of the parton evolution, such as $p_{\perp}$ ordering, can be examined using the Mercedes algorithm for jet analyses.

\section{Summary}\label{sec:summary}
The assumption of quark fragmentation universality implies that the current Breit hemisphere in DIS is expected to be very similar to a hemisphere in $e^+e^-$ annihilation. However, the experimental situations are different, and several corrections to universality are present. We have here proposed event cuts to improve the expected validity of quark fragmentation universality. DIS events with high-$p_\perp$ emissions, which have no correspondence in $e^+e^-$ events, are excluded, where the $p_\perp$-scale is reconstructed using jet cluster algorithms.

The effects of the cuts have been investigated using MC simulations, and the agreement between $e^+e^-$ and Breit frame results is shown to improve. The results after a cut in jet $p_\perp$ may depend on the choice of cluster algorithm.
We have investigated three different types of $k_\perp$ cluster schemes, and find our results to be  algorithm independent.

In the accepted low-$p_\perp$ DIS sample, heavy quarks are suppressed. This motivates a comparison with uds enriched $e^+e^-$ data, which are available from the experiments at the ${\mrm Z}^0$ pole, but not at lower energies. We have here presented a method, the ``thrust-cone algorithm'', to study scale evolutions of $e^+e^-$ quark hemispheres, using data from fixed energy experiments. With this algorithm, uds enriched data with high statistics from the LEP1 experiments can be compared to results for the current Breit hemisphere, over a large range of energies.

At high $Q^2$, the rapidity distance from the current Breit hemisphere to the remnant end is reduced. By comparing results from different MC programs, we find that the multiplicity in the current Breit hemisphere at very high $Q^2$ can be used to distinguish between different assumptions of remnant effects.

We also propose an observable which more directly investigates the properties of the target region. A current jet is defined with the ``Mercedes'' algorithm~\cite{jetscales}. The Mercedes current jet probes central rapidity regions in the Breit frame, but always stays on the current side of the highest $p_\perp$ emission of the event. This is a comparatively clean region, as compared to the forward target region. A corresponding jet definition can be studied in $e^+e^-$ experiments, which enables a comparison between the processes, and could facilitate the interpretation of the results. The properties of the Mercedes current jet is shown to be sensitive both to remnant effects and mechanisms behind the parton evolution, and can be used to discriminate between different existing models.

\subsection*{Acknowledgments}
I want to thank Leif L\"onnblad and G\"osta Gus\-taf\-son for their  significant contributions to this investigation.

This work was supported in part by the EU Fourth Framework Programme `Training and Mobility of Researchers',
Network `Quantum Chromodynamics and the Deep Structure of Elementary Particles', contract FMRX-CT98-0194 (DG 12 - MIHT).


\begin{thebibliography}{99}
\bibitem{empty}
  \bibl{K.H. Streng, T.F. Walsh, P.M. Zerwas}{Z. Phys.}{C2}{1979}{237}
\bibitem{QFUres}
  \bibl{ZEUS collaboration}{Z. Phys.}{C67}{1995}{93} \\
  \bibl{ZEUS collaboration}{Phys. Lett.}{B414}{1997}{428} \\
  \bibl{H1 collaboration}{Nucl. Phys.}{B504}{1997}{3} 
\bibitem{s_anomaly} 
  ZEUS preliminary results, contributed paper 809, ICHEP'98, Vancouver
\bibitem{jetalgs}
  \bibl{Yu.L. Dokshitzer, G.D. Leder, S. Moretti and B.R. Webber}{JHEP} {08}{1997}{001}\\
  For a review, see  \bibl{L. L\"onnblad, S. Moretti, T. Sj\"ostrand}{JHEP}{08}{1998}{001}
\bibitem{ktalg}
  \bibl{S. Catani, Yu.L. Dokshitzer, B.R. Webber}{Phys. Lett. }{B285}{1992}{291}
\bibitem{jetset}
  \bibl{M. Bengtsson, T. Sj\"ostrand}{Comp. Phys. Comm.}{39}{1986}{347}\\
  \bibl{T. Sj\"ostrand}{Comp. Phys. Comm.}{82}{1994}{74}
\bibitem{durham}
  \bibl{S. Catani, Yu.L. Dokshitzer, M. Olsson, G. Turnock and B.R. Webber}{Phys. Lett. }{B269}{1991}{432}
\bibitem{diclus}
  \bibl{L. L\"onnblad}{Z. Phys. }{C58}{1993}{471}
\bibitem{ariadne}
   \bibl{L. L\"onnblad}{Comp. Phys. Comm.}{71} {1992}{15}
\bibitem{lepto}
   \bibl{G. Ingelman, A. Edin, J. Rathsman}{Comp.Phys.Comm.}{101}{1997}{108} 
\bibitem{CDM}
  \bibl{G. Gustafson}{Phys. Lett. }{B175}{1986}{453}\\
  \bibl{G. Gustafson, U. Pettersson}{Nucl. Phys. }{B306}{1988}{746}\\
  \bibl{B. Andersson, G. Gustafson, L. L\"onnblad}{Nucl. Phys. }{B339}{1990}{393}
\bibitem{stringmodel}
  \bibl{B. Andersson, G. Gustafson, G. Ingelman, T. Sj\"ostrand}
	{Phys. Rep.} {97} {1983} {31}
\bibitem{charmrates}
  \bibl{H1 Collaboration}{Z.Phys}{C72}{1996}{593}\\
  H1 preliminary results, contributed paper 540, ICHEP'98, Vancouver\\
  ZEUS preliminary results, contributed paper 768, ICHEP'98, Vancouver
\bibitem{jetscales}
  \bibl{P. Ed\'en, G. Gustafson}{JHEP} {09}{1998}{015}
\bibitem{FGpuzzle}
  Yu.L. Dokshitzer, V.A. Khoze, A.H. Mueller and S.I. Troyan, Basics of Perturbative QCD, ({\it Editions Fronti\`eres, Gif-sur-Yvette, 1991}), p. 107-109
\bibitem{jung}
   H. Jung, L. J\"onsson, H. Kuster,  DESY preprint 98-051, hep-ph/9805396 v2
\bibitem{srm}
  \bibl{B. Andersson, G. Gustafson, L. L\"onnblad, U. Pettersson}{Z. Phys. }{C43}{1989}{625}
\bibitem{SCI}
   \bibl{G. Ingelman, A. Edin, J. Rathsman}{Phys. Lett.}{B336}{1996}{371}
\bibitem{hiQariadne}
   L. L\"onnblad, private communications, version of \textsc{Ariadne} to appear in future release.
\end{thebibliography}
\end{document}